\documentclass[reprint,aps,prc,amsmath,amssymb,showpacs,showkeys,twocolumns,floatfix]{revtex4-1}
\usepackage{dcolumn,graphicx,color,amsthm,enumerate,verbatim,subfigure}
\begin{document}

\preprint{APS/123-QED}

\title{Taylor approximation to treat nonlocality in scattering process}

\author{N. J. Upadhyay}
\email{neelam.upadhyay@cbs.ac.in}
\author{A. Bhagwat}
\email{ameeya@cbs.ac.in}
\affiliation{UM-DAE Centre for Excellence in Basic Sciences, Vidyanagari,
Mumbai-400098, India}
\date{\today}

\begin{abstract}
Study of scattering process in the nonlocal interaction framework leads
to an integro-differential equation. The purpose of the present work is 
to develop an efficient approach to solve this integro-differential equation
with high degree of precision. The method developed here employs Taylor
approximation for the radial wave function which converts the
integro-differential equation in to a readily solvable second-order
homogeneous differential equation. This scheme is found to be computationally 
efficient by a factor of 10 when compared to the iterative scheme developed 
in J.~Phys.~G~Nucl.~Part.~Phys.~{\bf 45},~015106~(2018). The calculated 
observables for neutron scattering off $^{24}$Mg, $^{40}$Ca, $^{100}$Mo 
and $^{208}$Pb with energies up to 10 MeV are found to be within at 
most 8$\%$ of those obtained with the iterative scheme. Further, we 
propose an improvement over the Taylor scheme that brings the 
observables so close to the results obtained by iterative scheme that 
they are visually indistinguishable. This is achieved without any 
appreciable change in the run time.
\end{abstract}


 
\maketitle

\section{Introduction}
The nonlocal interaction framework finds its application in diverse range
of scientific areas such as Physics and Quantum Biology 
\cite{phys1,phys2,phys3,phys4,bio1,wang16}. In such studies, the dynamics 
of the system is modelled in terms of integro-differential equation, 
which is usually difficult to solve analytically or even numerically. 
Hence, one has to resort to efficient techniques that yield highly precise 
solutions.

In the domain of nuclear physics, the many-body nature of the nucleus makes 
it imperative to study processes such as scattering and reaction in the 
nonlocal interaction framework \cite{lemere,balan97,frahn1,frahn2}. As a
consequence the conventional Schr\"{o}dinger equation becomes an
integro-differential equation, which is written as:
\begin{equation}
\left[\frac{\hbar^2}{2\mu}\nabla^2 + U_{SO}{\bf L}\cdot{\boldsymbol 
\sigma} + E \right]\Psi({\bf r}) = 
\int V({\bf r},{\bf r^\prime})\Psi({\bf r^\prime})d{\bf r^\prime}\,,
\label{eq1}
\end{equation}
where $U_{SO}\,{\bf L}\cdot{\boldsymbol \sigma}$ is the local spin-orbit
interaction, while $V({\bf r},{\bf r^\prime})$ is the nonlocal interaction 
kernel. Often this integro-differential equation is solved by using its
Fourier transform in momentum space, which leads to a Fredholm integral 
equation of the second kind. This approach has been used to study 
scattering and bound states of nuclei \cite{viv,so2013}.

Nevertheless, extensive studies in coordinate representation have been 
done to develop techniques that give precise solutions of Eq.(\ref{eq1}) 
\cite{frahn1,frahn2,pb,ali,ahmad,raw}. The most popular of them is the 
work of Perey and Buck \cite{pb} where the authors construct a local 
equivalent potential from the nonlocal nucleon-nucleus potential, which 
in turn is used to solve the integro-differential equation iteratively.

In our recent work \cite{nju18}, we have developed a readily implementable
technique using the second mean value theorem (MVT) of the integral 
calculus \cite{mvt} to solve the integro-differential equation. The 
advantage of the method is that it converts the integro-differential 
equation to the conventional Schr\"{o}dinger equation. However, as shown 
in \cite{nju18} to get a precise solution of Eq.(\ref{eq1}), an iterative 
scheme has been employed which is initiated by solution to the 
homogeneous equation. The iterative scheme, thus developed, is found to be 
robust but is time consuming due to its slow convergence rate.

In this paper we develop a very efficient technique to solve Eq.(\ref{eq1})
that yields results with precision comparable to those obtained by 
the full iterative MVT (IMVT) scheme of Ref.~\cite{nju18}. For this 
purpose, we use Taylor approximation for the radial wave function which
has been known for a long time, see for example Ref.~\cite{glend}.
This method converts the integro-differential equation to a homogeneous 
second-order differential equation that can be easily solved. Further, 
to test the accuracy of the technique we have studied neutron scattering 
off different targets spanning the entire periodic table in the energy 
range up to 10 MeV.

The Taylor approximation approach developed to solve Eq.(\ref{eq1}) 
forms the subject matter of Section-II. Results along with discussions 
are presented in Section-III, while the conclusions are given in 
Section-IV.

\section{Formalism}
In order to study scattering of neutrons from the spin-zero nucleus, we 
start with partial wave expansion of Eq.(\ref{eq1}). This is done by 
writing scattering wave function, $\Psi({\bf r})$, and the nonlocal 
interaction kernel, $V({\bf r},{\bf r^\prime})$ as:
\begin{eqnarray}
\nonumber
\Psi({\bf r})&=&\sum_{l m_l m_s} \frac{u_{jl}(r)}{r}\,
\bigg\langle l\,\frac{1}{2}\,m_l\,m_s\,\bigg|\,j\,(m_l+m_s) \bigg\rangle
\\
&&\hspace{1.5cm}\times\,\,i^l\,Y_{lm_l}(\Omega_r)\,\chi_{\frac{1}{2} m_s},
\label{eq2}
\\
V({\bf r},{\bf r^\prime})&=&\sum_{l} \frac{(2l+1)}{4\pi}
\frac{g_{l}(r,r^\prime)}{rr^\prime}\,P_l\left({\rm cos}\,\theta\right)\,,
\label{eq3}
\end{eqnarray}
with $\theta$ being the angle between ${\bf r}$ and ${\bf r^\prime}$
\cite{pb,nju18}. 

The resulting radial equation is
\begin{eqnarray}
&&\hspace{1.1cm} \hat{\mathcal{L}}\,u_{jl}(r)\,=\,\frac{2\mu}{\hbar^2}
\int_0^{\infty} g_l(r,r^\prime) u_{jl}(r^\prime) dr^\prime,
\label{eq4}
\\
\nonumber
&&{\rm where,}\,\,\hat{\mathcal{L}} \equiv \left[\frac{d^2}{dr^2} - 
\frac{l(l+1)}{r^2} + \frac{2\mu U_{SO}(r)}{\hbar^2}
f_{jl} + \frac{2\mu\,E}{\hbar^2} \right]\,,
\\
\nonumber
&&f_{jl} = \frac{1}{2}\left[j(j+1)-l(l+1)- \frac{3}{4}\right]\,\,\,
{\rm with}\,\, j = l\pm1/2\,,\,\,{\rm and}
\\
&&g_l(r,r^\prime)=2\pi r r^\prime\,
\int_{-1}^{1} V \left({\bf r},{\bf r^\prime}\right)\,
P_l\left({\rm cos}\,\theta\right) d\left({\rm cos}\,\theta\right)\,.
\label{eq5}
\end{eqnarray}
For the interaction kernel, $V \left({\bf r},{\bf r^\prime}\right)$, 
we use Frahn and Lemmer prescription \cite{frahn1,frahn2},
\begin{equation}
V({\bf r},{\bf r^\prime})= \frac{1}{\pi^{3/2}\beta^3}\,
{\rm exp}\left[-\,\frac{\left|{\bf r} - {\bf r^\prime}\right|^2}{\beta^2}
\right] U \left(\frac{\left|{\bf r} + {\bf r^\prime}\right|}{2}\right)\,,
\label{eq6}
\end{equation}
where $\beta$ is the nonlocal range parameter. In this work, the energy 
and mass independent nucleon-nucleus potential, $U$, is taken to be of 
Wood-Saxon form. The parameters for this potential are taken from 
Tian {\it et al.} \cite{tpm15}. For further details regarding potential
refer to Sec.~2.1 of \cite{nju18}. Following the convention adopted in 
\cite{nju18}, this potential will be referred to as `TPM15'.

To begin with, we enlist the salient features of the IMVT approach 
developed earlier in \cite{nju18}.

\subsection{The IMVT Approach}
In the IMVT approach, using the second mean value theorem of the integral
calculus \cite{mvt}, the nonlocal interaction kernel is written as
\begin{equation}
\int_0^{\infty} g_l(r,r^\prime) u_{jl}(r^\prime) dr^\prime\,
\approx\,u_{jl}(r) \int_0^{\infty} g_l(r,r^\prime) dr^\prime,
\end{equation}
where the observation that $g_l(r,r^\prime)$ is strongly peaked at
$r$=$r^\prime$ is incorporated. Substituting this in 
Eq.(\ref{eq4}), we obtain a homogeneous equation of the form
\begin{equation}
\hat{\mathcal{L}}\,u_{jl}(r)\,=\,\frac{2\mu U^{\rm eff}_l(r)}{\hbar^2} 
u_{jl}(r),
\label{homo}
\end{equation}
where the dominant effect of nonlocality is contained in the effective 
local potential, $\displaystyle{U^{\rm eff}_l(r) = \int_0^{\infty} 
g_l(r,r^\prime) dr^\prime}$. Further, this potential is independent of 
energy but depends upon partial waves. 

The solution of Eq.(\ref{eq4}) is obtained by implementing an iterative 
scheme. This scheme is initiated by the solution to the above homogeneous 
equation (Eq.(\ref{homo})) and the subsequent iterants are obtained by 
solving:
\begin{eqnarray}
\hat{\mathcal{L}}u^{i+1}_{jl}(r) &-& \frac{2\mu\,U^{\rm eff}_l(r)}
{\hbar^2} u^{i+1}_{jl}(r)
\\
\nonumber
&&\hspace{-1cm}=\frac{2\mu}{\hbar^2}\int_0^{\infty} g_l(r,r^\prime)
u^i_{jl}(r^\prime)\,dr^\prime -\frac{2\mu U^{\rm eff}_l(r)}{\hbar^2} 
u^i_{jl}(r)\,,
\end{eqnarray}
for all $i \ge 0$. The iterations are continued till the 
absolute value of difference between the logarithmic derivatives of 
the wave functions at the matching radius in the $i^{\rm th}$ and the 
$(i+1)^{\rm th}$ steps match within the desired precision, $\epsilon$.

The scattering wave function is obtained with the radial step size of
0.02 fm and matching radius of 20 fm. To obtain converged logarithmic
derivative with $\epsilon \sim 10^{-6}$ at a given energy, the 
typical run time required is about an hour on a single Intel i7-6700 
processor. Further, the run time scales almost linearly with the 
number of partial waves and energy, making the method time consuming.
The fact that the IMVT scheme though robust, is time consuming, limits
its usability to routine and large scale calculations.

To partially remedy this limitation, in Ref.\cite{nju18} it was proposed
that instead of full iterative procedure computation can be done with
only one iteration. This results in speed-up of calculations by a factor 
of 4 as compared to the IMVT scheme. However, we would like to point out 
that the success of this solution depends strongly on the choice of 
nucleon-nucleus potential, mass of target as well as the projectile energy. 
For example, in case of neutron scattering off $^{208}$Pb and energies up 
to 2 MeV, it was found that the results for TPM15 potential with one 
iteration deviates from the IMVT results by as much as 20$\%$. Hence, 
it is important to develop a robust and efficient scheme to obtain 
precise solution to Eq.(\ref{eq4}).

\subsection{The Taylor Approximation Approach}
\begin{figure*}[htp!]
\centering
\subfigure[][~Real part]{
\centering
\includegraphics[scale=0.40]{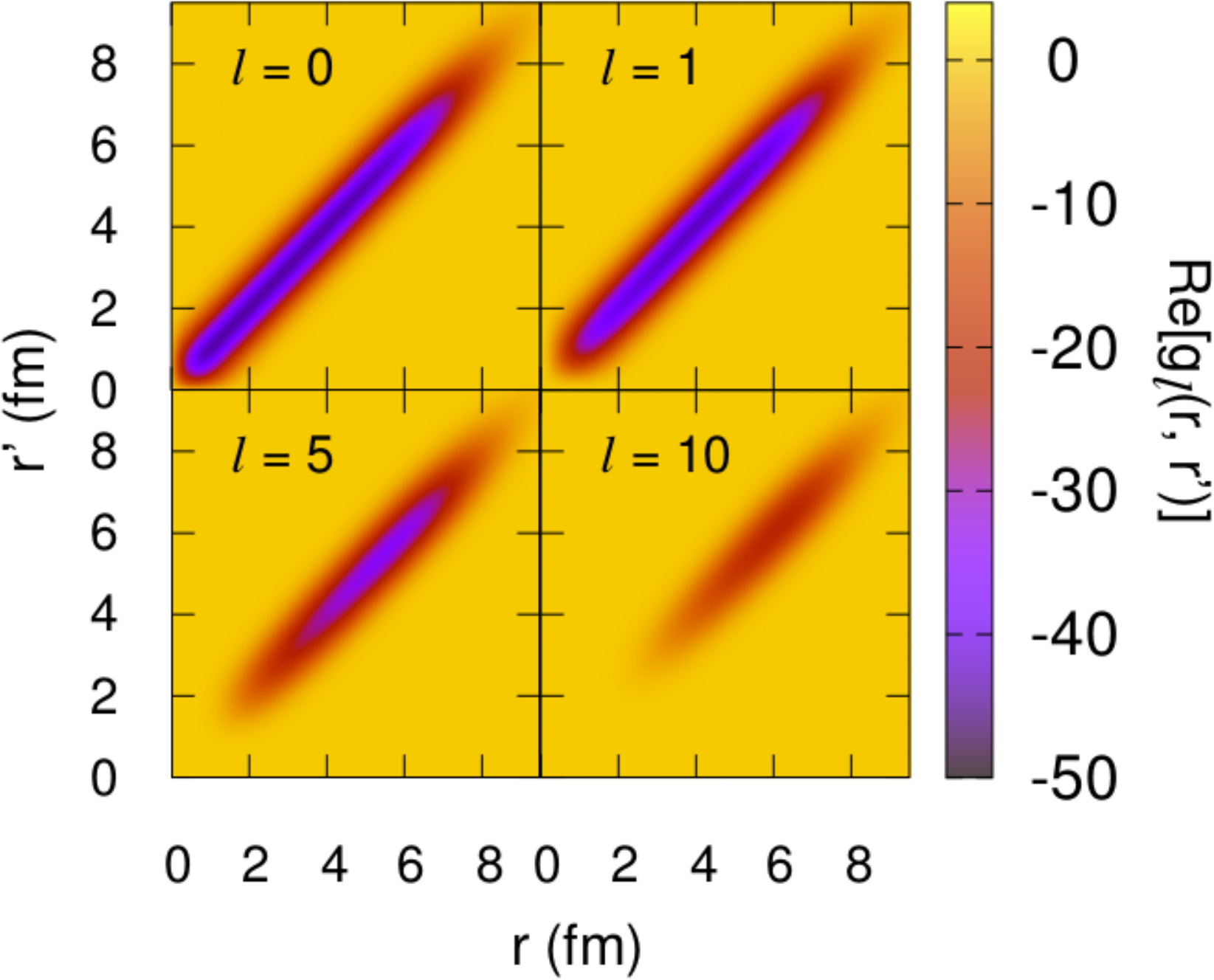}}
\quad\quad
\subfigure[][~Imaginary part]{
\centering
\includegraphics[scale=0.40]{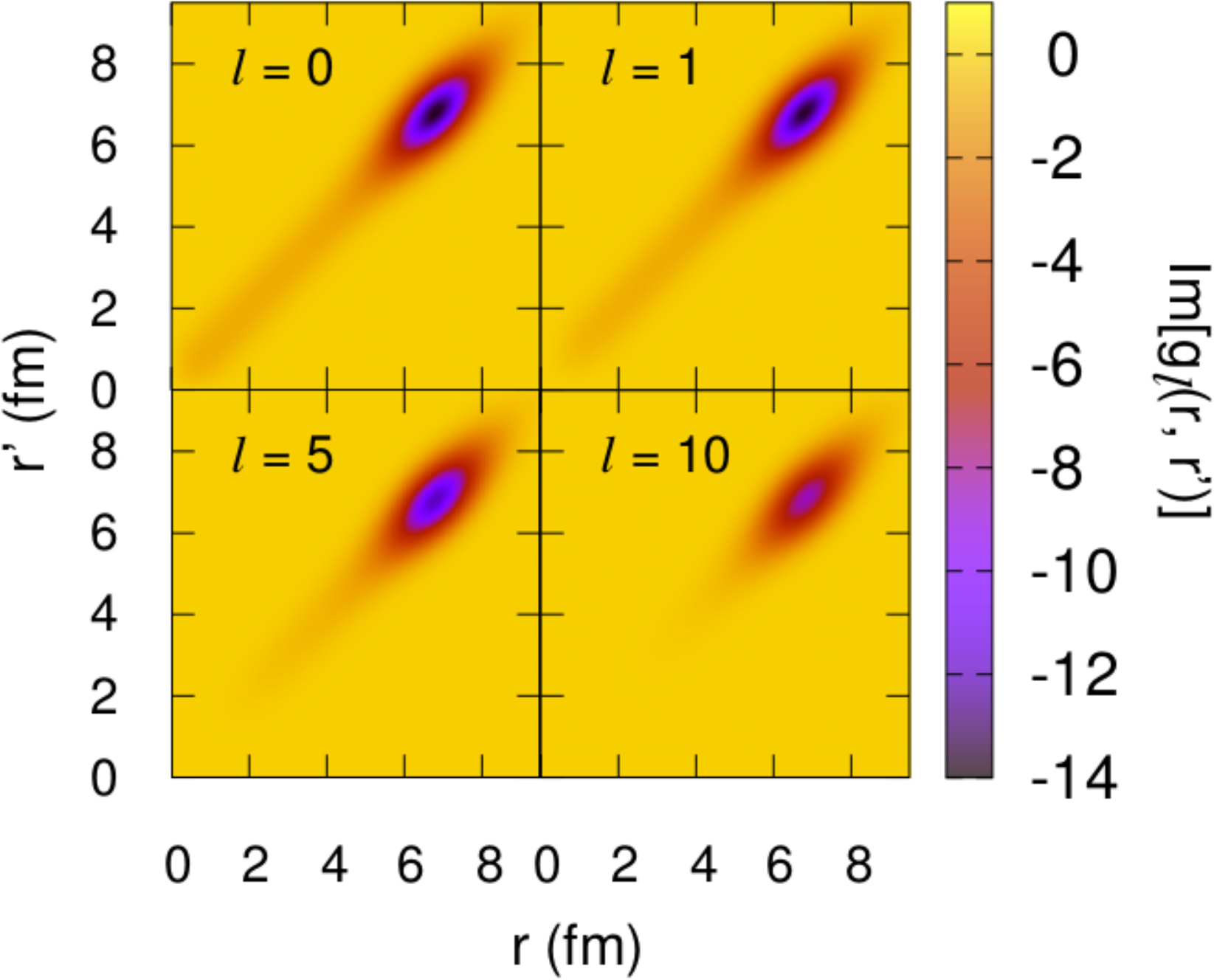}}
\caption{Behavior of $g_l(r,r^\prime)$ as a function of distance for 
different $l$. Calculations are done for neutron scattering off 
$^{208}$Pb using TPM15 potential \cite{tpm15}.}
\label{f1}
\end{figure*}

The principal objective of this work is to devise an efficient method 
to solve Eq.(\ref{eq4}) with precision comparable to that obtained by
the IMVT approach. To achieve this we examine the structure of nonlocal
kernel, $g_l(r,r^\prime)$, closely. As an illustration, in Fig.~\ref{f1}
we show the nonlocal kernel for neutron scattering off $^{208}$Pb using
TPM15 potential \cite{tpm15}. As can be seen from the figure, the 
nonlocality is dominant around the line $r$=$r^\prime$. Any appreciable 
deviation from this line makes the contribution from the nonlocal kernel 
insignificant. 

Motivated by this observation, we write $r^\prime$=$r+\Delta$ and expand 
the wave function $u_{jl}(r^\prime)$ about $r$=$r^\prime$ using Taylor's
theorem \cite{kline} as
\begin{equation}
u_{jl}(r^\prime) = P_n(r^\prime) + R_n(r^\prime)\,,\,\,\,\,(n\,\ge\,0)
\end{equation}
where $P_n(r^\prime)$ is the $n$-th order Taylor polynomial, written as
\begin{equation}
P_n(r^\prime) = u_{jl}(r) + \sum_{\lambda=1}^n 
\frac{\Delta^\lambda}{\lambda!}\,\frac{d^\lambda u_{jl}(r)}{dr^{\lambda}}\,,
\end{equation}
while the remainder term, $R_n(r^\prime)$, is written as
\begin{equation}
R_n(r^\prime) = \frac{\Delta^{n+1}}{(n+1)!}\,
\frac{d^{n+1}u_{jl}(\xi)}{dr^{n+1}}\,,
\end{equation}
for some $\xi$ between $r$ and $r^\prime$. 
Since the wave functions are guaranteed to be differentiable up to 
second-order for nonsingular potentials, we expand $P_n(r^\prime)$ up to 
first-order ($n$=1) and retain the remainder term giving
\begin{equation}
u_{jl}(r^\prime) = u_{jl}(r) + \Delta\,\frac{du_{jl}(r)}{dr} +
\frac{\Delta^2}{2}\,\frac{d^2u_{jl}(\xi)}{dr^2}\,.
\label{expand}
\end{equation}
As the kernel is sharply peaked around $r$=$r^\prime$ (see Fig.~\ref{f1}),
we take $\xi \approx r$. Thus, the integral on the right hand side of 
Eq.(\ref{eq4}) can be written as:
\begin{eqnarray}
\nonumber
&&\frac{2\mu}{\hbar^2}\int_0^{\infty}g_l(r,r^\prime) u_{jl}(r^\prime) 
dr^\prime =  u_{jl}(r) I_{l0}(r) + \frac{du_{jl}(r)}{dr} I_{l1}(r)
\\
&&\hspace{5.0cm} + \frac{d^2u_{jl}(r)}{dr^2} I_{l2}(r)
\label{expand2}
\\
&&\hspace{0.2cm}{\rm where,}\,\,\,I_{ln}(r) = \frac{2\mu}{\hbar^2}
\int_0^{\infty} \frac{\Delta^n}{n!}\,g_l(r,r^\prime)dr^\prime\,,
\end{eqnarray}
with $0 \le n \le2$. Substituting this in Eq.(\ref{eq4}) and rearranging 
the terms, we get a homogeneous second-order differential equation 
written as
\begin{equation}
\label{neweq}
\hat{\mathcal{O}}u_{jl}(r) = 0\,,
\end{equation}
where,
\begin{eqnarray}
&&\hat{\mathcal{O}} \equiv \frac{d^2}{dr^2}-X_l(r) 
\frac{d}{dr} + W_l(r) \bigg[-\frac{l(l+1)}{r^2}\,\,\,\,\,\,\, 
\\
\nonumber
&&\hspace{2cm}+\,\frac{2\mu U_{SO}(r)}{\hbar^2} f_{jl}+ 
\frac{2\mu\,E}{\hbar^2} - I_{l0}(r)\bigg]\,,
\\
&&\hspace{-0.5cm}
X_l(r) = \frac{I_{l1}(r)}{(1-I_{l2}(r))} \,\,\,{\rm and}\,
\,\,W_l(r) = \frac{1}{(1-I_{l2}(r))}\,.
\end{eqnarray}
The obtained equation is a simple second order differential equation 
that can be readily solved. The first-order derivative appears 
explicitly in Eq.(\ref{neweq}) and enough care has to be taken to 
evaluate it accurately. For this we revisit the behaviour of the 
wave function near the origin. 

Near the origin, Eq.(\ref{neweq}) becomes
\begin{equation}
\left[\frac{d^2}{dr^2}-\frac{l(l+1)}{r^2}+\frac{2\mu\,E}{\hbar^2}\right]
u_{jl}(r)\,\approx\,0\,,\,\,\,\,\,\,\,\,(r \to 0)\,.
\end{equation}
Redefining $\displaystyle{u_{jl}(r)\,=\,r^{l+1}\,\phi_l(r)}$, we get
\begin{equation}
\phi^{\prime\prime}_l(r) + \frac{2(l+1)}{r}\phi^\prime_l(r) + 
\frac{2\mu\,E}{\hbar^2} \phi_l(r) \,\approx\,0\,.
\end{equation}
To solve the above differential equation, we use Frobenius method \cite{frob}
and obtain
\begin{equation}
\phi_l(r \to 0) \, = \, \sum_{\lambda = 0}^\infty 
\frac{(-)^l \, (kr)^{2\lambda}}{2^\lambda\, \lambda! \, (2l+2\lambda+1)!!}\,\,.
\end{equation}
Retaining the first four term of the series (up to $\lambda$=3), the expression
for $u_{jl}(r)$ near origin is written as
\begin{widetext}
\begin{equation}
u_{jl}(r \to 0) \,\approx\,\frac{r^{l+1}}{(2l+1)!!} \left[ 1 - 
\frac{r^2k^2}{2(2l+3)} + \frac{r^4k^4}{8(2l+3)(2l+5)} - 
\frac{r^6k^6}{48 (2l+3)(2l+5)(2l+7)} \right].
\label{origin}
\end{equation} 
\end{widetext}
Now the first-order derivative appearing in Eq.(\ref{neweq}) can be 
calculated accurately using Eq.(\ref{origin}). This expression also 
complies with the fact that $u_{j0}(0)$=0, $u^\prime_{j0}(0)$=1 for 
$l$=0 and $u_{jl}(0)$=$u^\prime_{jl}(0)$=0 for $l \ne$0. Finally, 
using Eq.(\ref{origin}) and its derivative as the initial conditions, 
we solve Eq.(\ref{neweq}) using the fourth order Runge-Kutta method 
\cite{rk4}.

\section{Results}
\subsection{The Taylor Approximation Approach}
To illustrate the method developed above, we consider neutron scattering 
off $^{24}$Mg, $^{40}$Ca, $^{100}$Mo and $^{208}$Pb with energies up to 
10 MeV. Calculations are done with TPM15 potential \cite{tpm15}.
Similar to the IMVT calculations, the radial step size is taken to be 0.02~fm,
while the matching radius is assumed to be 20~fm.

\begin{figure}[htp!]
\centering
\subfigure[][]{
\centering
\includegraphics[scale=0.35]{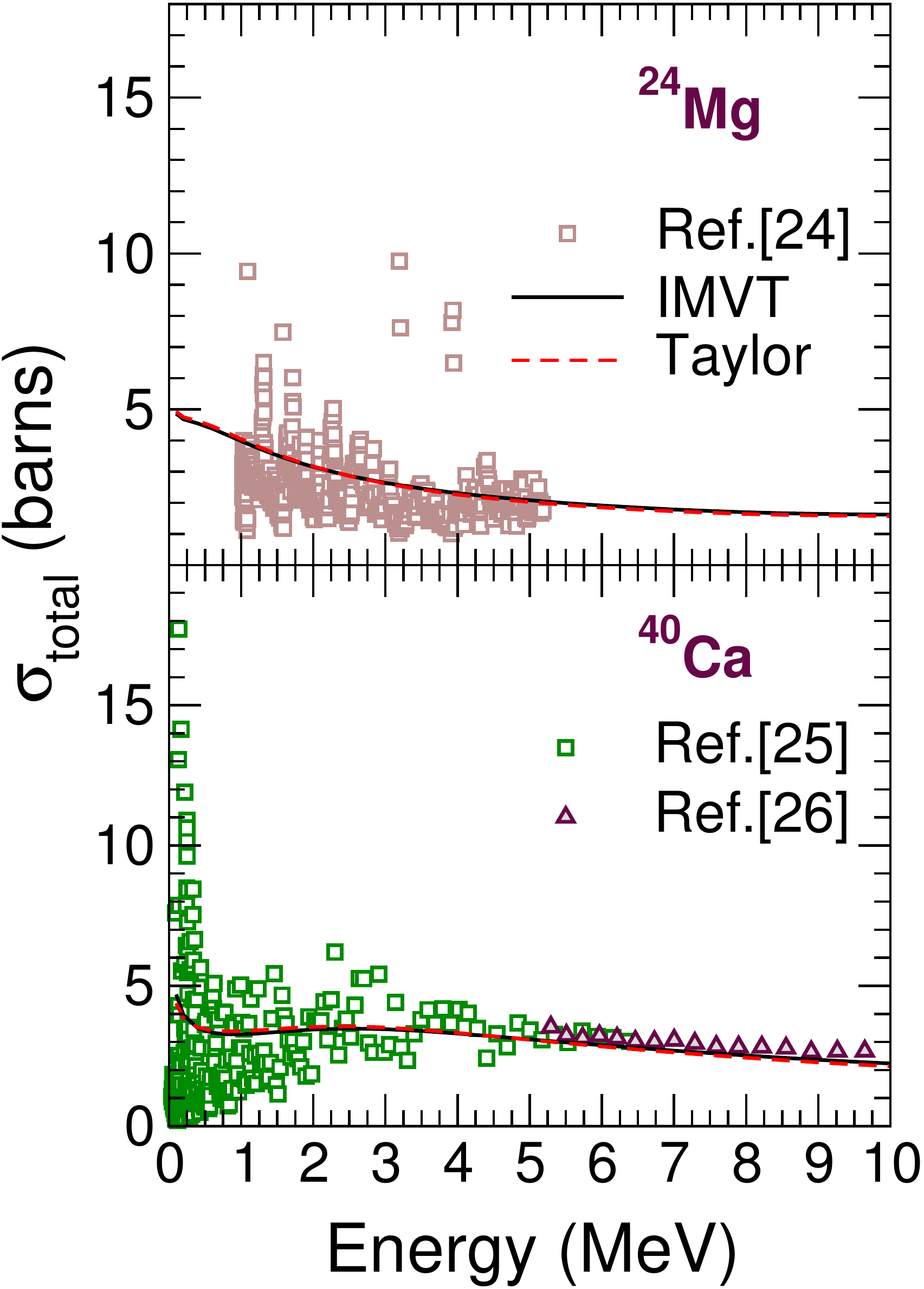}}
\quad\quad
\subfigure[][]{
\centering
\includegraphics[scale=0.35]{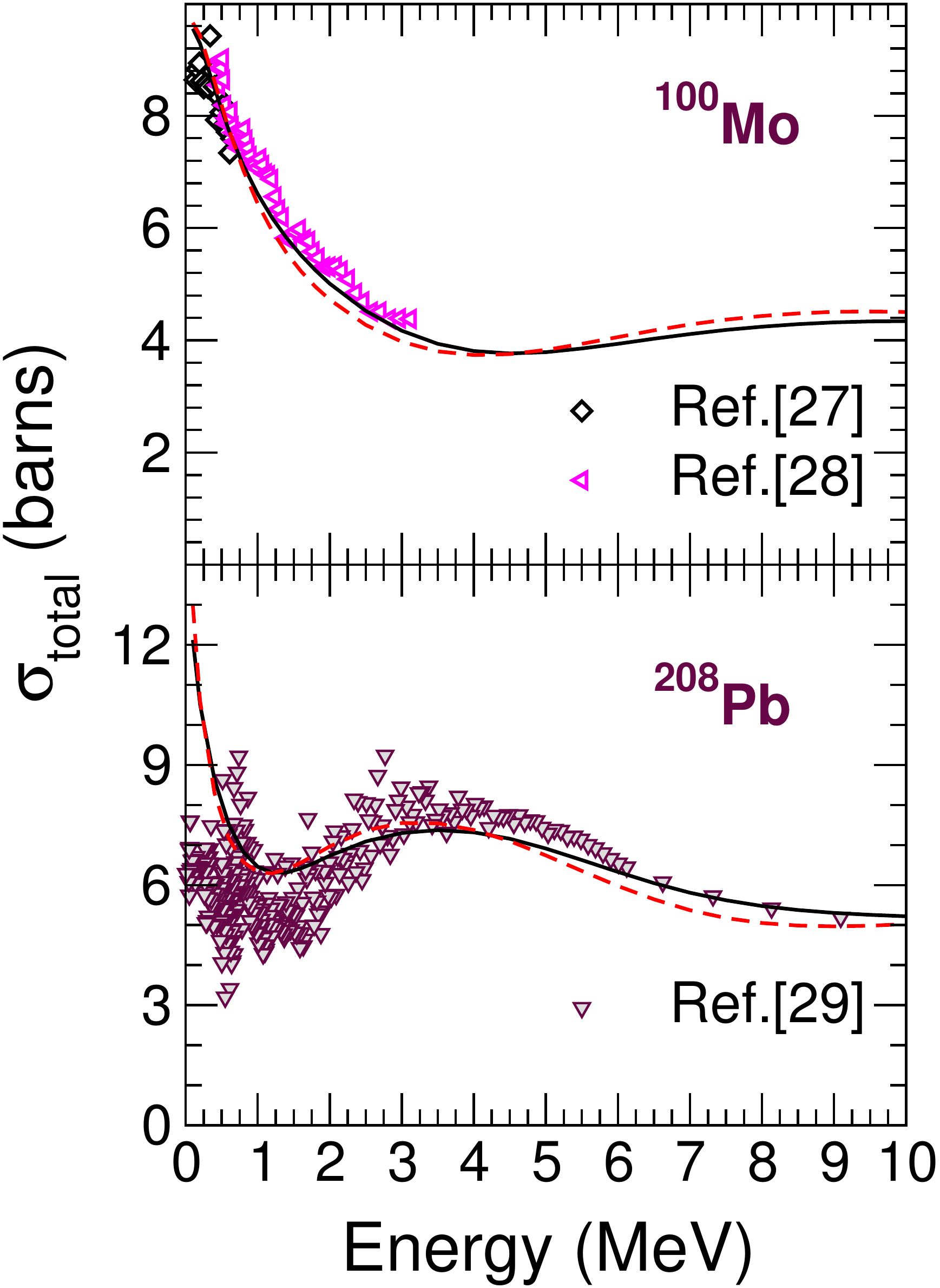}}
\caption{Calculated total cross sections for neutron scattering off 
$^{24}$Mg, $^{40}$Ca, $^{100}$Mo and $^{208}$Pb along with the data 
\cite{bomm,fowler,ca93,diva,pas,harvey}. The Taylor results are shown
by the dashed red line while the IMVT results \cite{nju18} are shown by 
the solid black line. Calculations are done using TPM15 potential 
\cite{tpm15}.}
\label{f2}
\end{figure}

In order to test the accuracy of the Taylor scheme, in Fig.~\ref{f2}
we compare the results of the present work (labeled as Taylor) with 
those obtained by the IMVT scheme (labeled as IMVT) along with the data 
\cite{bomm,fowler,ca93,diva,pas,harvey}. The cross sections calculated 
using the Taylor scheme are found to be close to those obtained by the 
IMVT scheme. Further, both the calculated results are in good agreement 
with the experiments.

\begin{figure}[ht!]
\centering{
\includegraphics[scale=0.45]{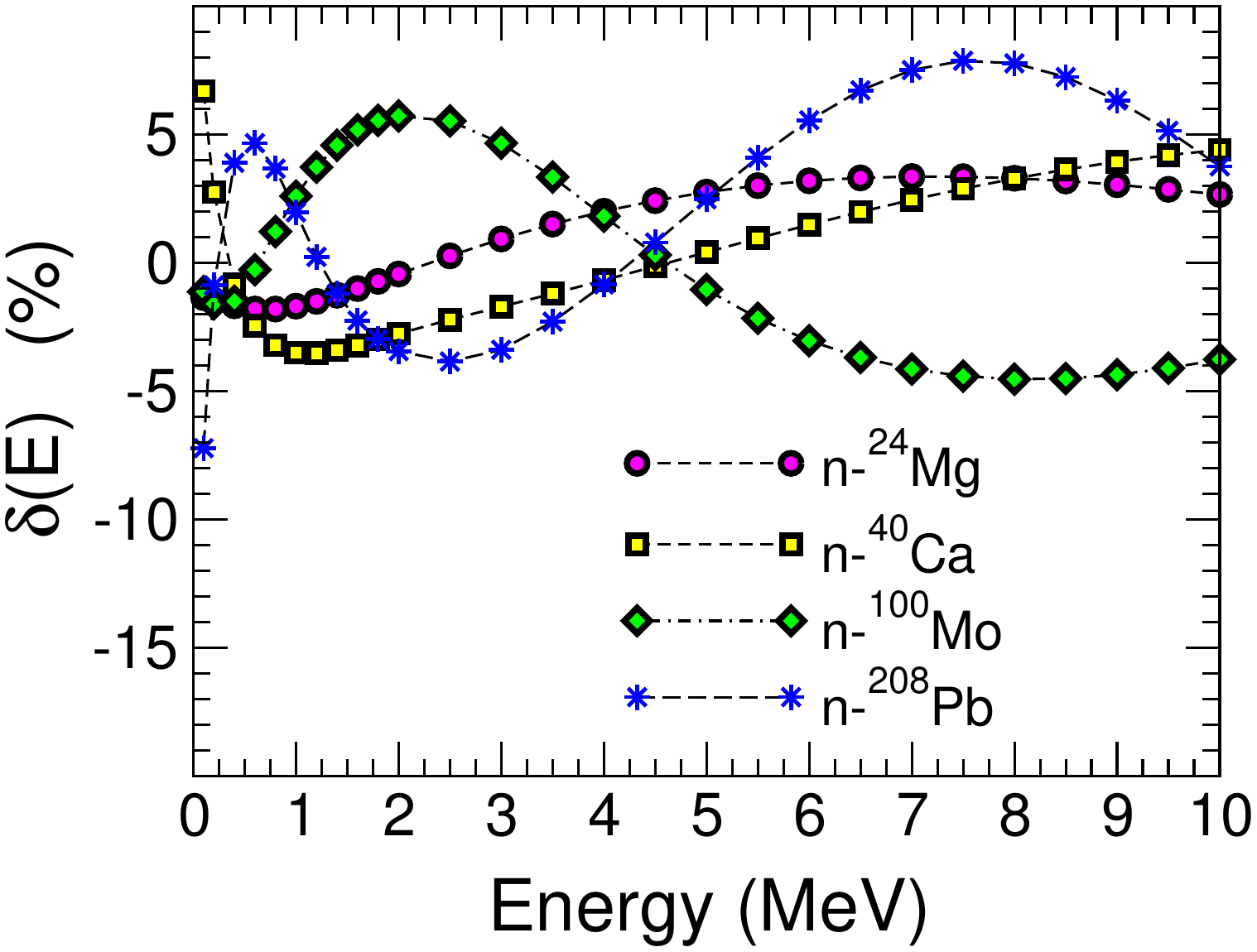}
\caption{Quantity $\delta(E)$ as a function of energy, 
$E$, for neutron scattering off different nuclei.}
\label{f3}}
\end{figure}

At a finer level, the Taylor and the IMVT results slightly differ from each 
other. This difference can be quantified by studying the behavior of 
$\delta(E)$, defined as
\begin{equation}
\delta(E)\,=\,\frac{\sigma_{\rm IMVT}(E)-\sigma_{\rm Taylor}(E)}
{\sigma_{\rm IMVT}(E)}\times 100
\end{equation}
with respect to neutron energy, $E$. In Fig.~\ref{f3} we plot the quantity 
$\delta(E)$ as a function of  energy for all the targets. It is seen that 
the cross sections obtained by the Taylor scheme are within at the most 8$\%$ 
of those obtained by the IMVT scheme for all the cases. 

The typical run time required for the Taylor scheme is about 5 minutes for 
a given energy on a single Intel i7-6700 processor. This demonstrates 
that the Taylor scheme is computationally efficient by a factor of 10 in 
comparison to IMVT approach and at the same time yields results within 
8$\%$ of the IMVT results, which is gratifying.

\subsection{Iterative Perturbation Approach}
The Taylor scheme devised in the previous section can be improved further 
without any appreciable change in the run time. This is achieved by solving
Eq.(\ref{eq4}) using an Iterative Perturbation approach (IPA). In this 
approach, the exact solution is expressed as a perturbation series
\begin{equation}
u_{jl}(r)\,=\,u_{jl}^0(r)\,+\,\sum_{k=1}^\infty u_{jl}^k(r)\,,
\end{equation}
where $u_{jl}^0(r)$ is the solution of Eq.(\ref{neweq}) and 
$u_{jl}^k(r)$ is the higher-order correction that quantifies the 
deviation from the exact solution. These higher-order corrections
are obtained with the help of following iterative scheme:
\begin{equation}
\hat{\mathcal{O}} u_{jl}^{i+1}(r) - W_l(r)\,\xi_{jl}^i(r)\,= 0\,,
\end{equation}
where,
\begin{equation}
\xi_{jl}^i(r)\,=\,\frac{2\mu}{\hbar^2}
\int_0^{\infty} g_l(r,r^\prime) u_{jl}^i(r^\prime) dr^\prime\,-\,
\hat{\mathcal{G}}\,u_{jl}^i(r)\,,
\end{equation}
where $i\ge 0$ and $\displaystyle{\hat{\mathcal{G}}\,\equiv\,I_{l0}(r) + I_{l1}(r) 
\frac{d}{dr}+I_{l2}(r) \frac{d^2}{dr^2}}$. The corrected wave function, 
$u_{jl}(r)$, thus obtained, is then matched with the free state wave 
function to calculate the $S$-matrix, which in turn is employed in 
computation of observables.

\begin{figure}[ht!]
\centering{
\includegraphics[scale=0.45]{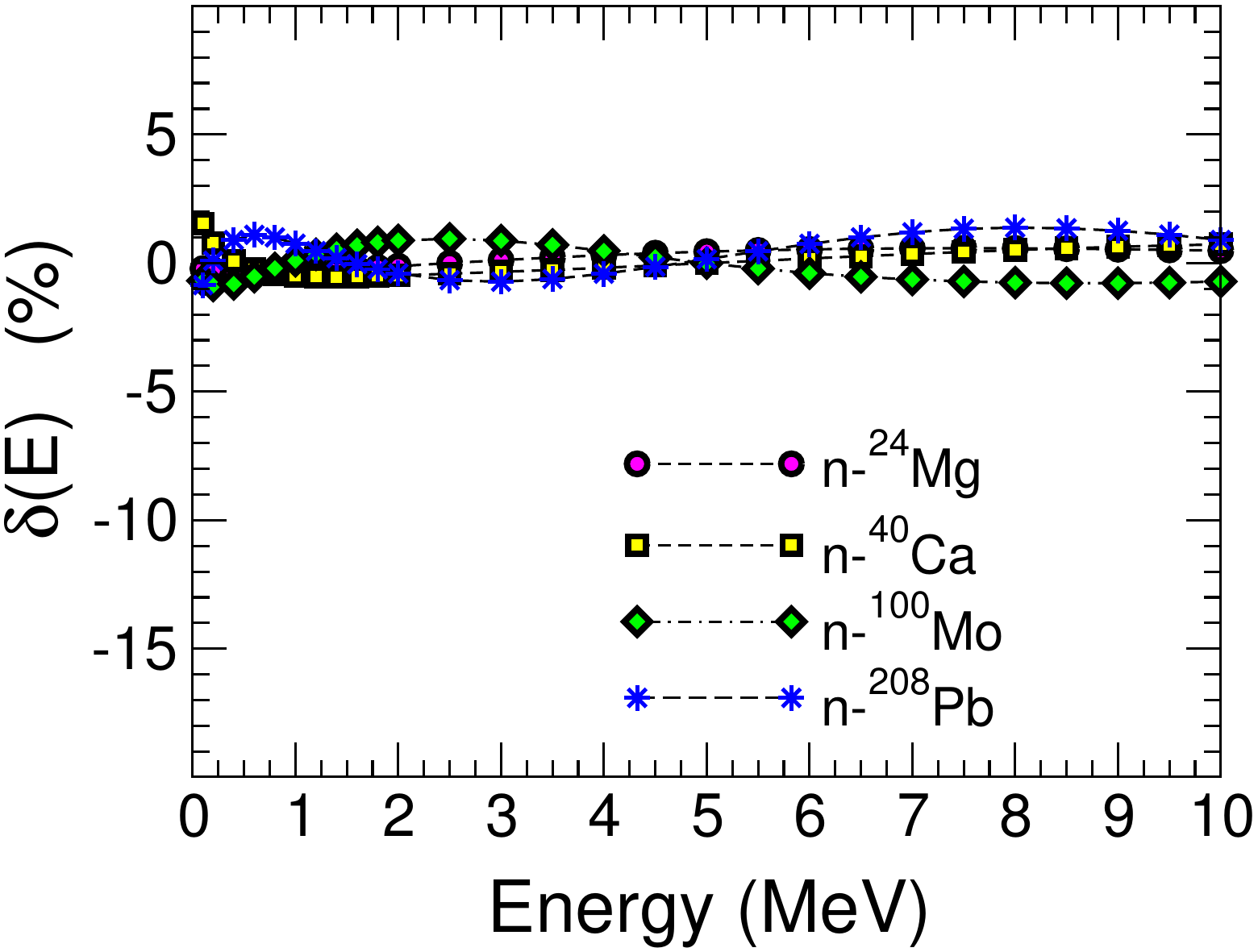}
\caption{Same as Fig~\ref{f3}, but $\delta(E)$ computed for IPA5
relative to IMVT.}
\label{f4}}
\end{figure}

In Fig.~\ref{f4} we quantify the accuracy of IPA cross sections calculated 
after 5 iterations (referred as IPA5) relative to the IMVT cross sections 
by plotting $\delta(E)$ as a function of energy. The IPA5 cross sections
are found to be within 2$\%$ of the IMVT cross sections at all energies 
for all the cases. 

\begin{figure}[htb!]
\centering
\subfigure[][]{
\centering
\includegraphics[scale=0.35]{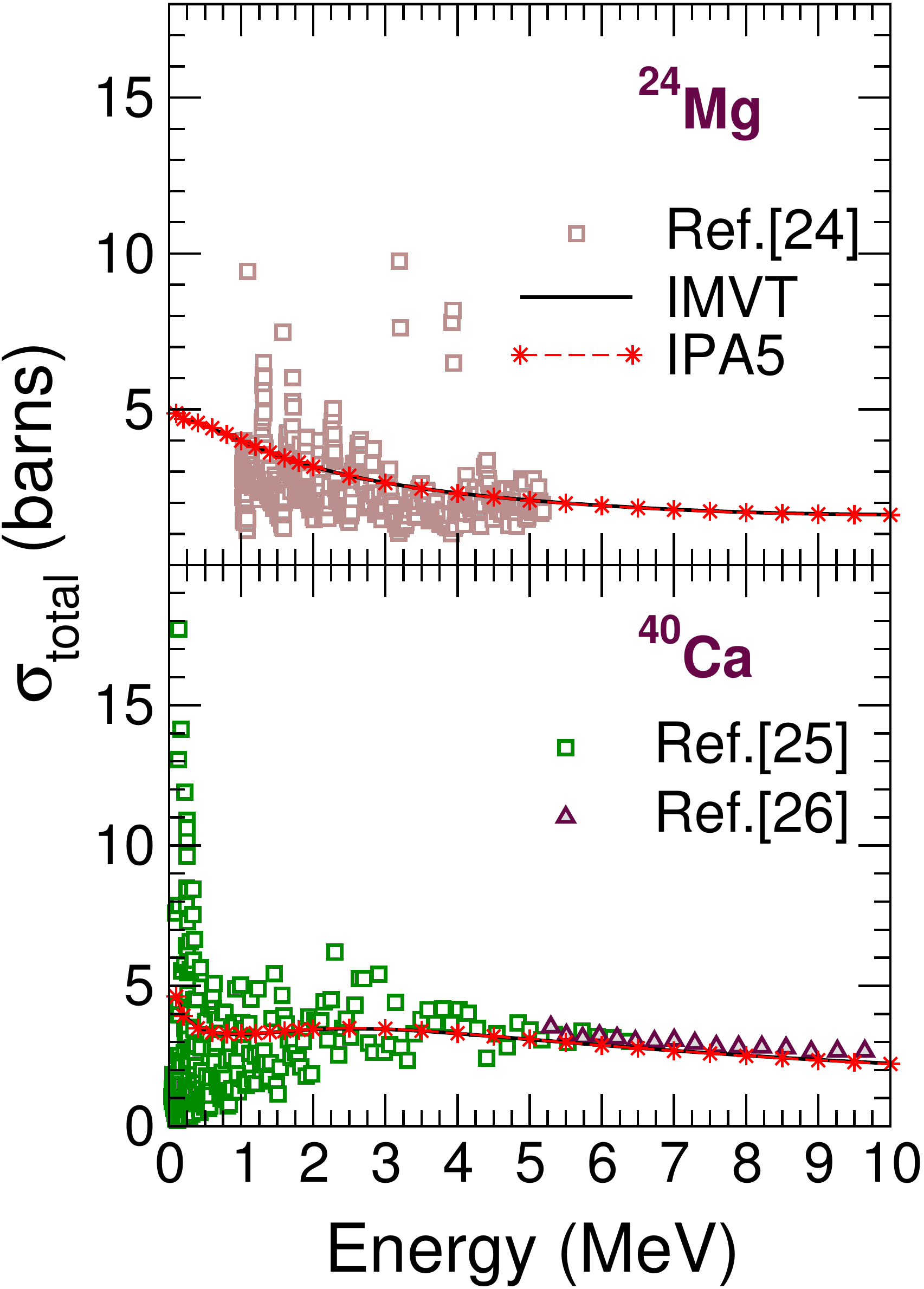}}
\quad\quad
\subfigure[][]{
\centering
\includegraphics[scale=0.35]{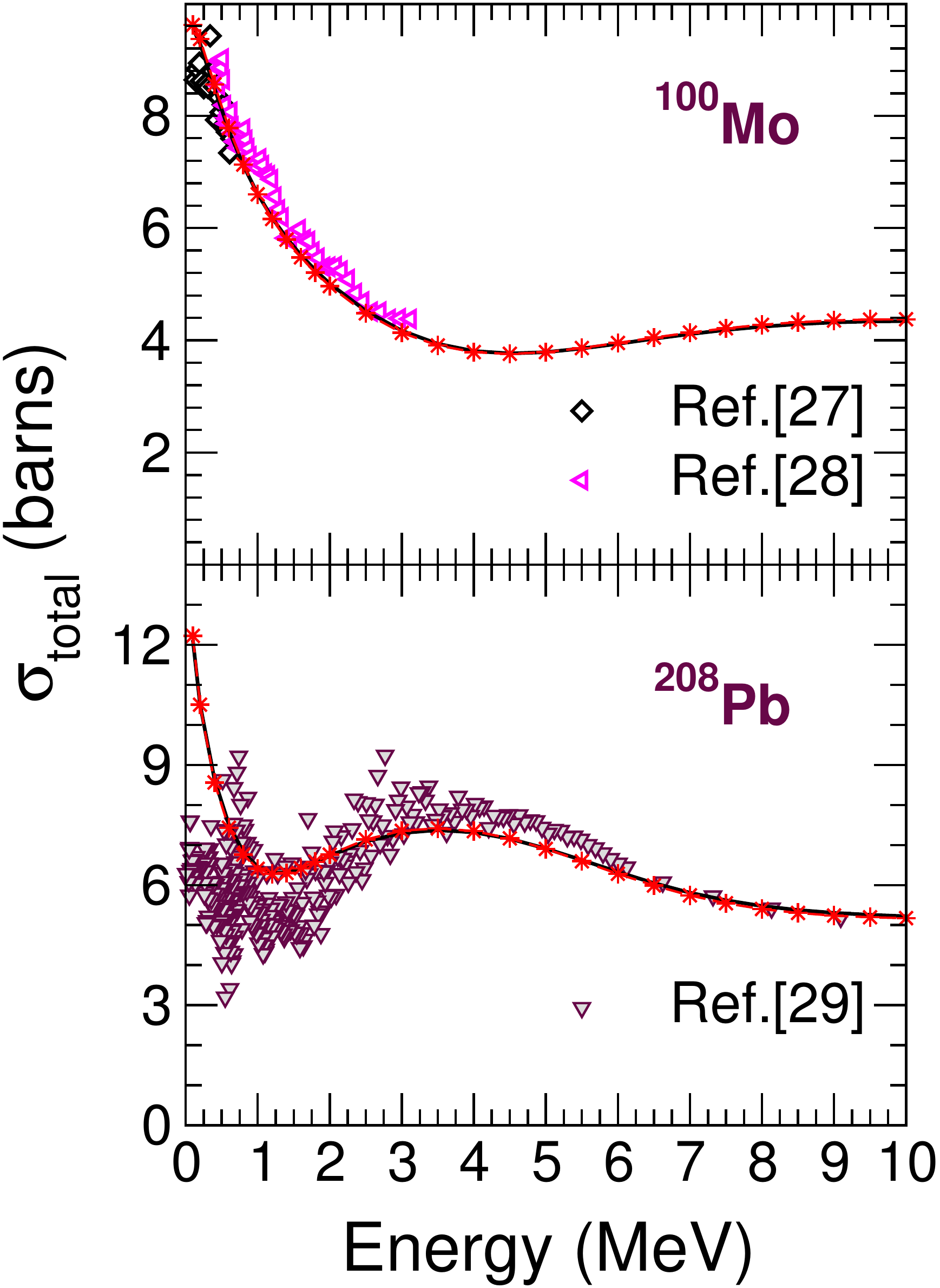}}
\caption{Same as Fig~\ref{f2}, but for IPA5. The IPA5 results are shown 
by the starred red line.}
\label{f5}
\end{figure}

Further, in Fig.~\ref{f5} we show the total cross sections calculated by 
IPA5 for neutron scattering off different nuclei along with the data 
\cite{bomm,fowler,ca93,diva,pas,harvey}. Visually, the IPA5 and the IMVT 
results are indistinguishable. Computationally there is no significant 
change at all in the run time when compared with that required for the 
Taylor scheme. 

These results demonstrate that the improved technique IPA yields highly
precise solution to Eq.(\ref{eq4}). Further, the technique is highly 
efficient since we have achieved a speed-up by a factor of 10 as compared
to the IMVT scheme, which is extremely significant in particular when it 
comes to large scale computations.

\begin{figure*}[ht!]
\centering
\subfigure[][~n-$^{24}$Mg scattering]{
\centering
\includegraphics[scale=0.34]{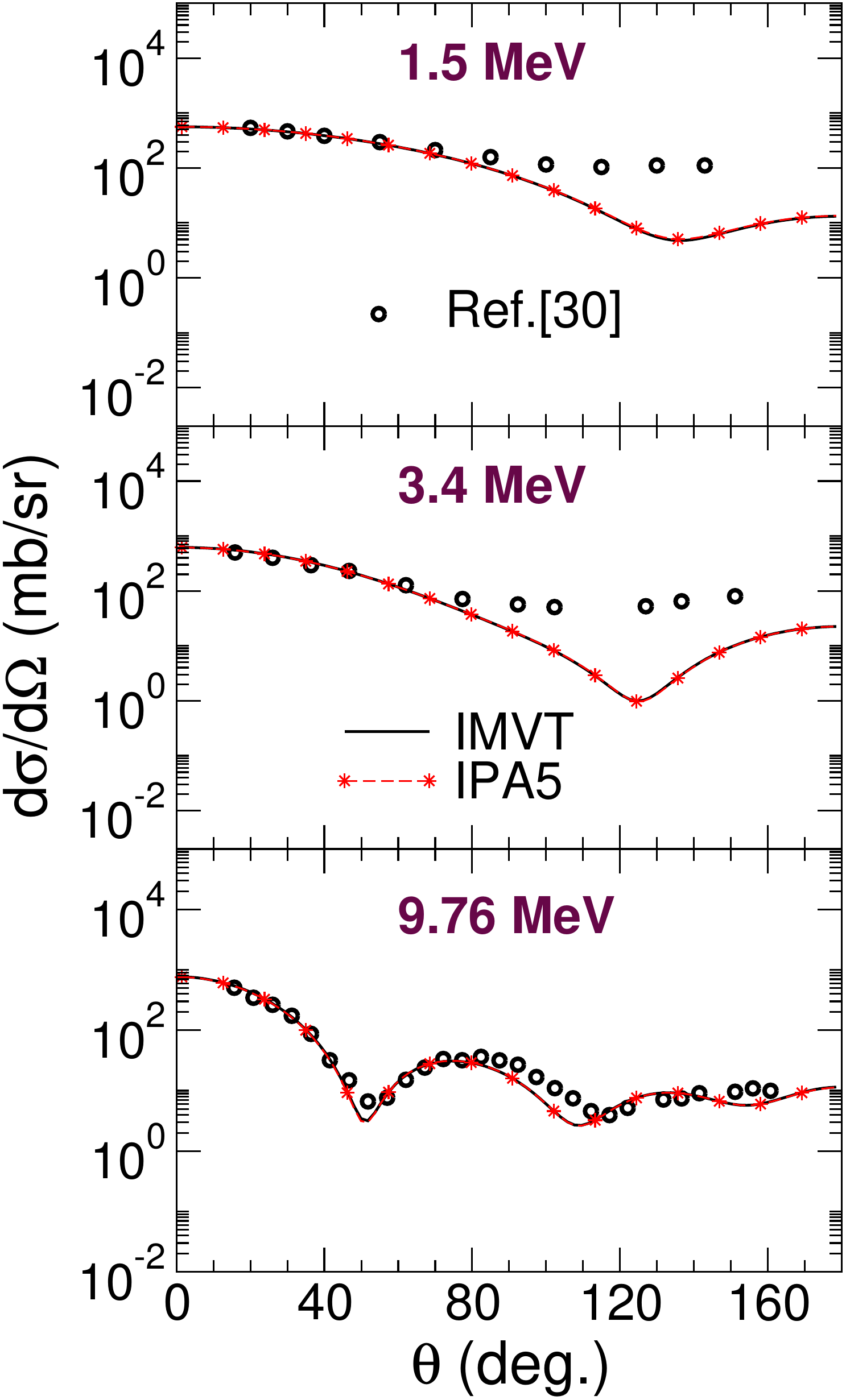}}
\quad\quad
\subfigure[][~n-$^{40}$Ca scattering]{
\centering
\includegraphics[scale=0.34]{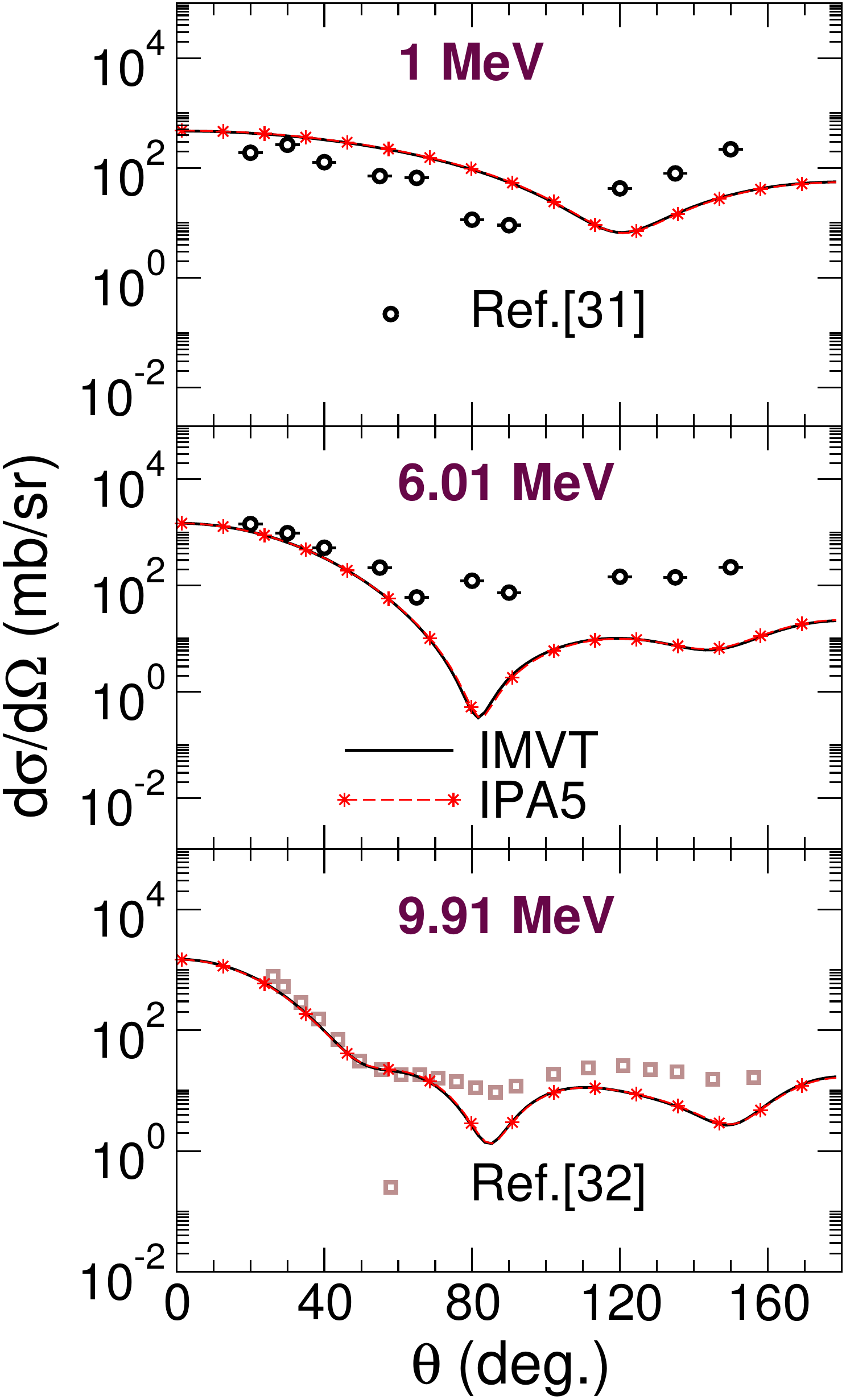}}
\caption{Calculated angular distributions for neutron scattering off
$^{24}$Mg and $^{40}$Ca along with the data \cite{ang24mg,toep,torn}. 
The IPA5 results are shown by the starred red line while the IMVT 
results \cite{nju18} are shown by the solid black line. Calculations 
are done using TPM15 potential \cite{tpm15}.}
\label{f6}
\end{figure*}

\begin{figure*}[ht!]
\centering
\subfigure[][~n-$^{100}$Mo scattering]{
\centering
\includegraphics[scale=0.34]{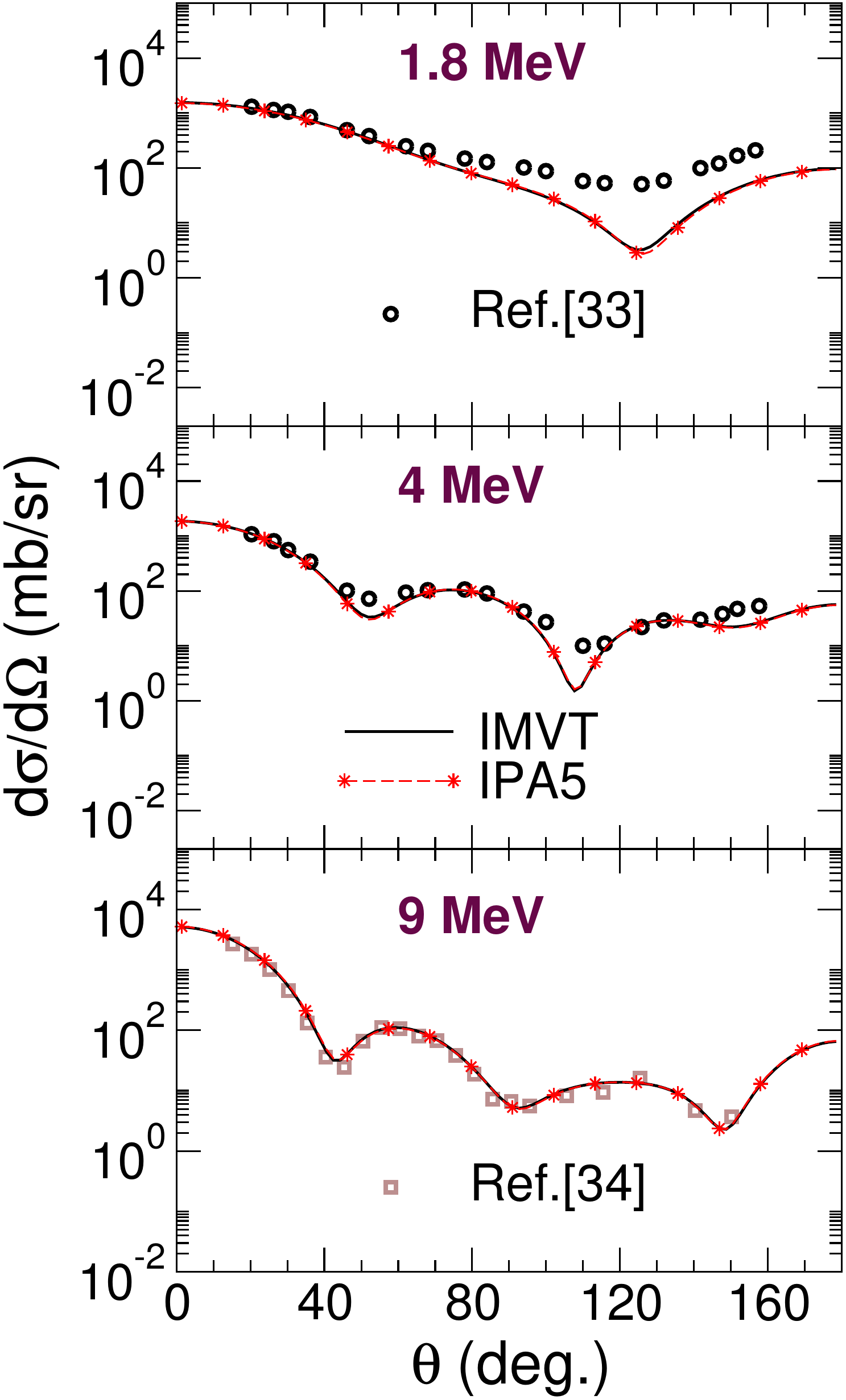}}
\quad\quad
\subfigure[][~n-$^{208}$Pb scattering]{
\centering
\includegraphics[scale=0.34]{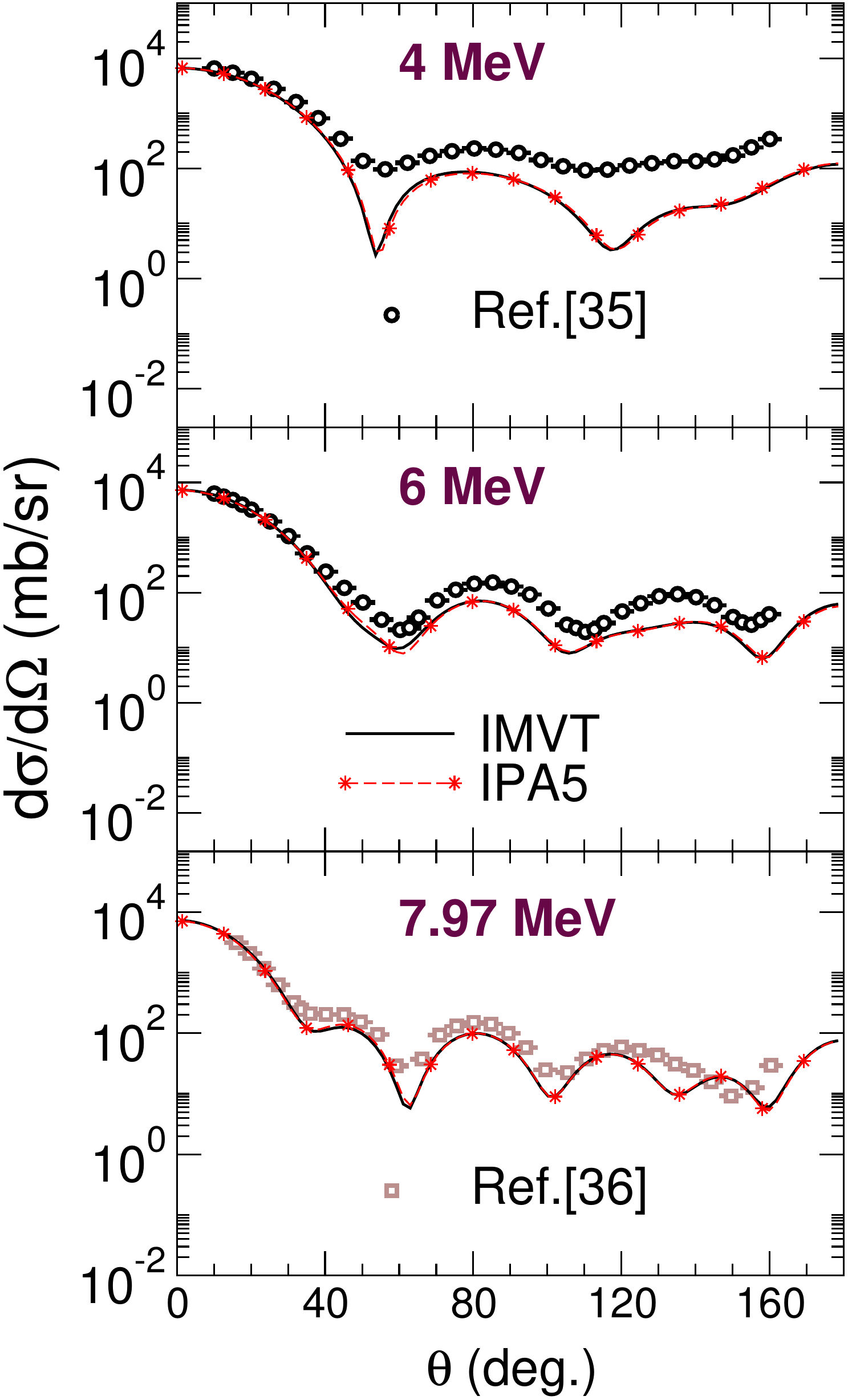}}
\caption{Same as Fig.~\ref{f6}, but for neutron scattering off 
$^{100}$Mo and $^{208}$Pb along with the data \cite{smith,rapa,annand,rob}.}
\label{f7}
\end{figure*}

\subsection{Angular Distributions}

For completeness, in Figs.~\ref{f6}-\ref{f7} we show various calculated 
angular distributions along with the experimental data 
\cite{ang24mg,toep,torn,smith,rapa,annand,rob}. As observed earlier, 
again the IPA5 and the IMVT results are found to be indistinguishable. For 
$^{24}$Mg and $^{40}$Ca we observe that the calculated results are reasonably 
consistent with the data at low energies, while those for $^{100}$Mo 
and $^{208}$Pb are in good accord at all the energies. It may be mentioned 
that the parameters for TPM15 potential are obtained by fitting the nucleon 
scattering data on nuclei ranging from $^{27}$Al to $^{208}$Pb with 
incident energies around 10 MeV to 30 MeV. Probably a better agreement can
be achieved with more appropriate choice of potential. Further investigations
along these lines are in progress.

\subsection{Robustness of IPA}
In the present work, a separable form for the interaction kernel 
(see Eq.(6)) is used, which is given as
\begin{equation}
V({\bf r},{\bf r^\prime})= H\left(\left|{\bf r} - {\bf r^\prime}\right|
\right) U \left(\frac{\left|{\bf r} + {\bf r^\prime}\right|}{2}\right)\,.
\end{equation}
The function $H\left(\left|{\bf r} - {\bf r^\prime}\right|\right)$ is 
chosen to be a Gaussian with the range $\beta$=0.9~fm (as given in 
\cite{tpm15}) and is normalized to unity. To establish the robustness 
of IPA, it is essential to study its sensitivity to different forms 
of nonlocality.

\subsubsection{Impact of different forms of nonlocality}
As a first step, we explore the impact of different forms of 
$H\left(\left|{\bf r} - {\bf r^\prime}\right|\right)$ with same 
normalization and rms radius but different shapes. For this we consider 
an exponential function:
\begin{equation}
H\left(\left|{\bf r} - {\bf r^\prime}\right|\right)\,=\,
\frac{1}{8 \pi \alpha^3}\,{\rm exp}\left(-\frac{|\bf{r}-\bf{r^\prime}|}
{\alpha}\right)\,.
\end{equation}
which is normalized similar to the Gaussian function. Further, the 
nonlocal range $\alpha$ has been chosen in such a way that both the
Gaussian and exponential form factors have the same rms radii, 
giving: $\alpha\,=\,\beta/\sqrt{8}$. 

\begin{figure}[htb!]
\centering
\subfigure[][]{
\centering
\includegraphics[scale=0.35]{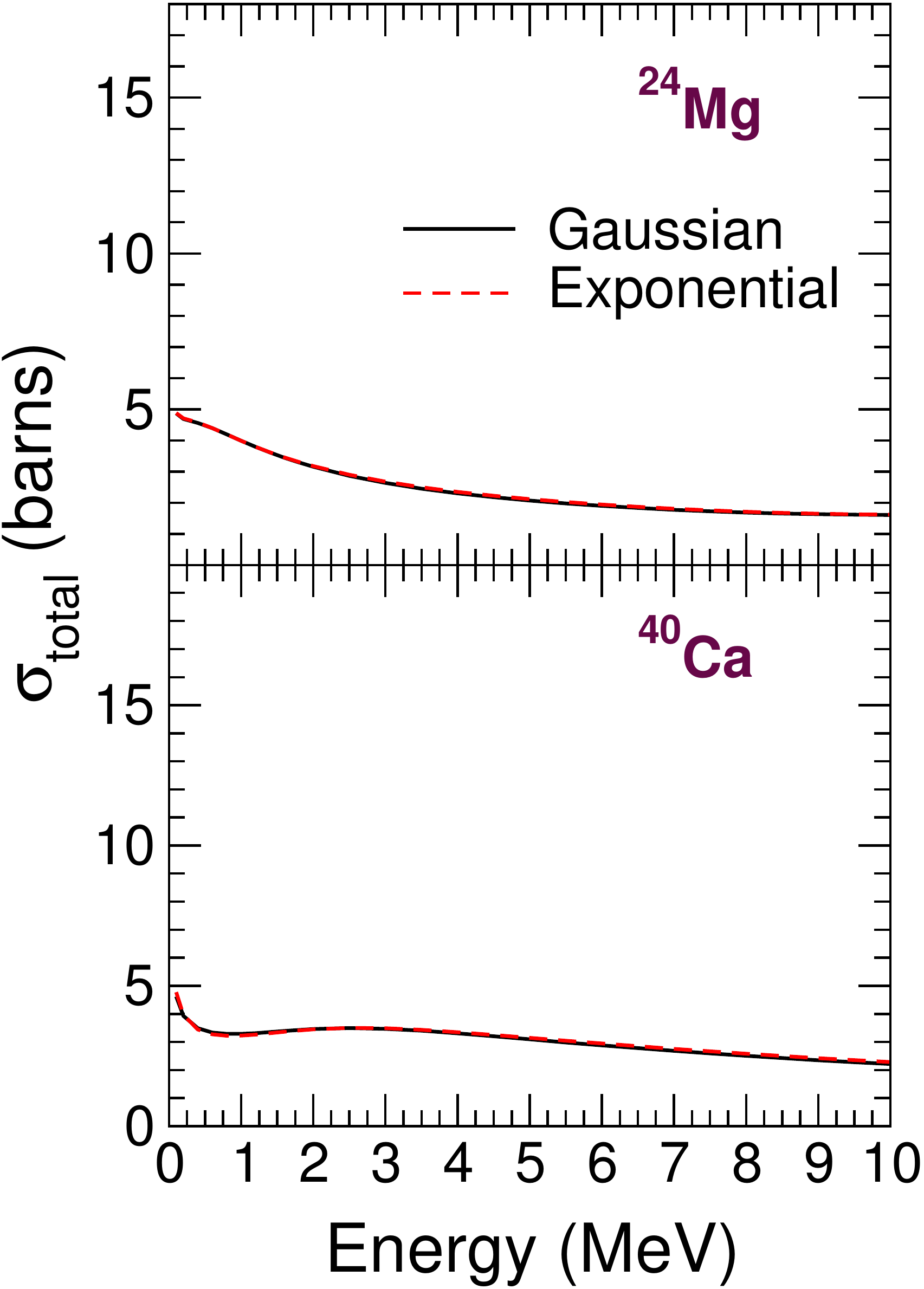}}
\quad\quad
\subfigure[][]{
\centering
\includegraphics[scale=0.35]{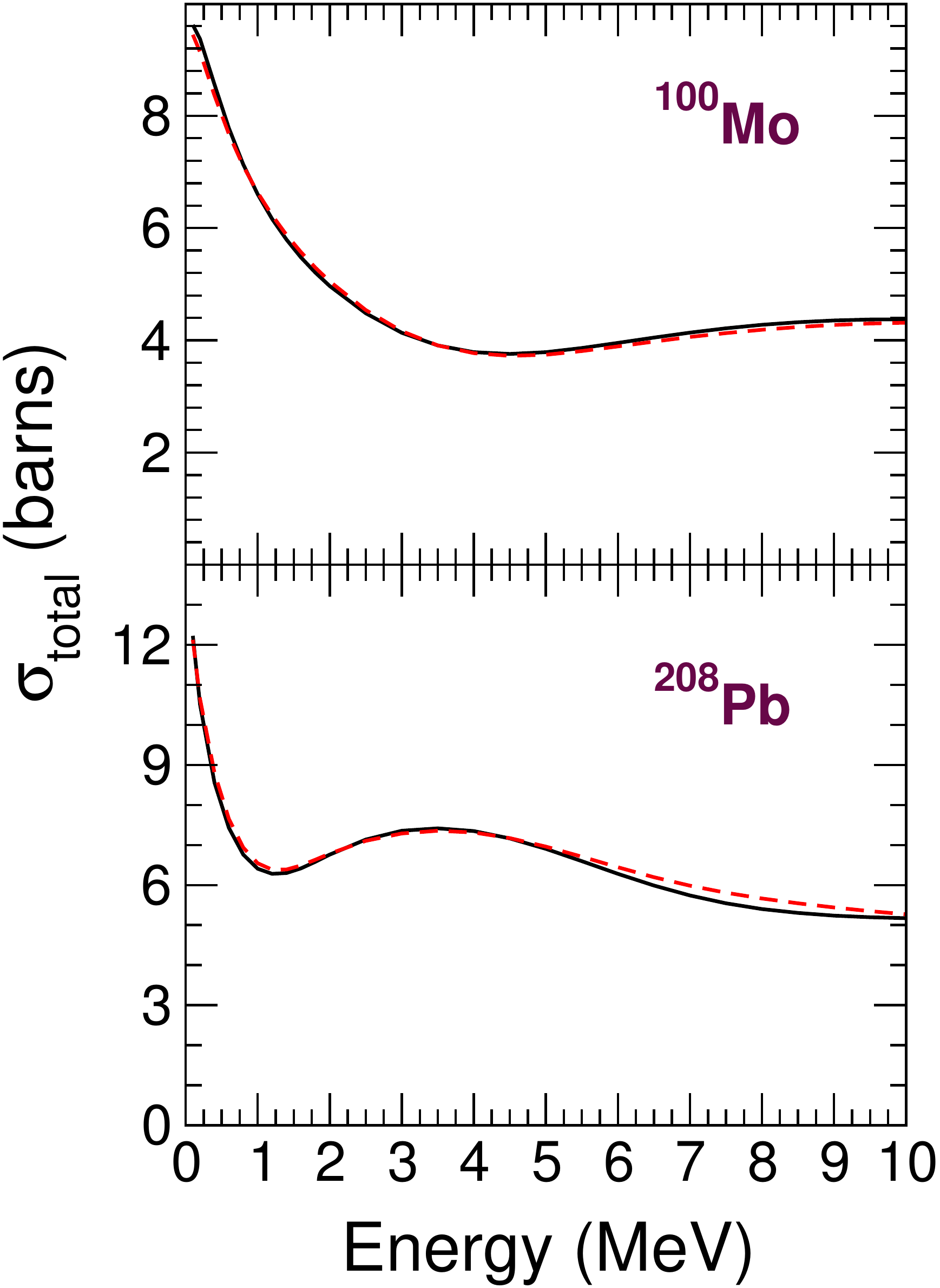}}
\caption{Calculated total cross sections for neutron scattering off
$^{24}$Mg, $^{40}$Ca, $^{100}$Mo and $^{208}$Pb. The results for
Gaussian form of nonlocality are shown by the solid black line while 
the results for exponential form are shown by the dashed red line. 
Calculations are done by IPA5 using TPM15 potential \cite{tpm15}. For 
exponential form, $\alpha$=0.318~fm.}
\label{f8}
\end{figure}

In Fig.~\ref{f8} we show the total cross sections calculated by using
Gaussian and exponential forms of nonlocality in IPA5 calculations 
for neutron scattering off different nuclei. As expected, different 
forms for $H\left(\left|{\bf r} - {\bf r^\prime}\right|\right)$ having 
same normalization and rms radius give similar results \cite{nju18}.

\subsubsection{Impact of different ranges of nonlocality}
\begin{figure*}
\centering{
\includegraphics[scale=0.5]{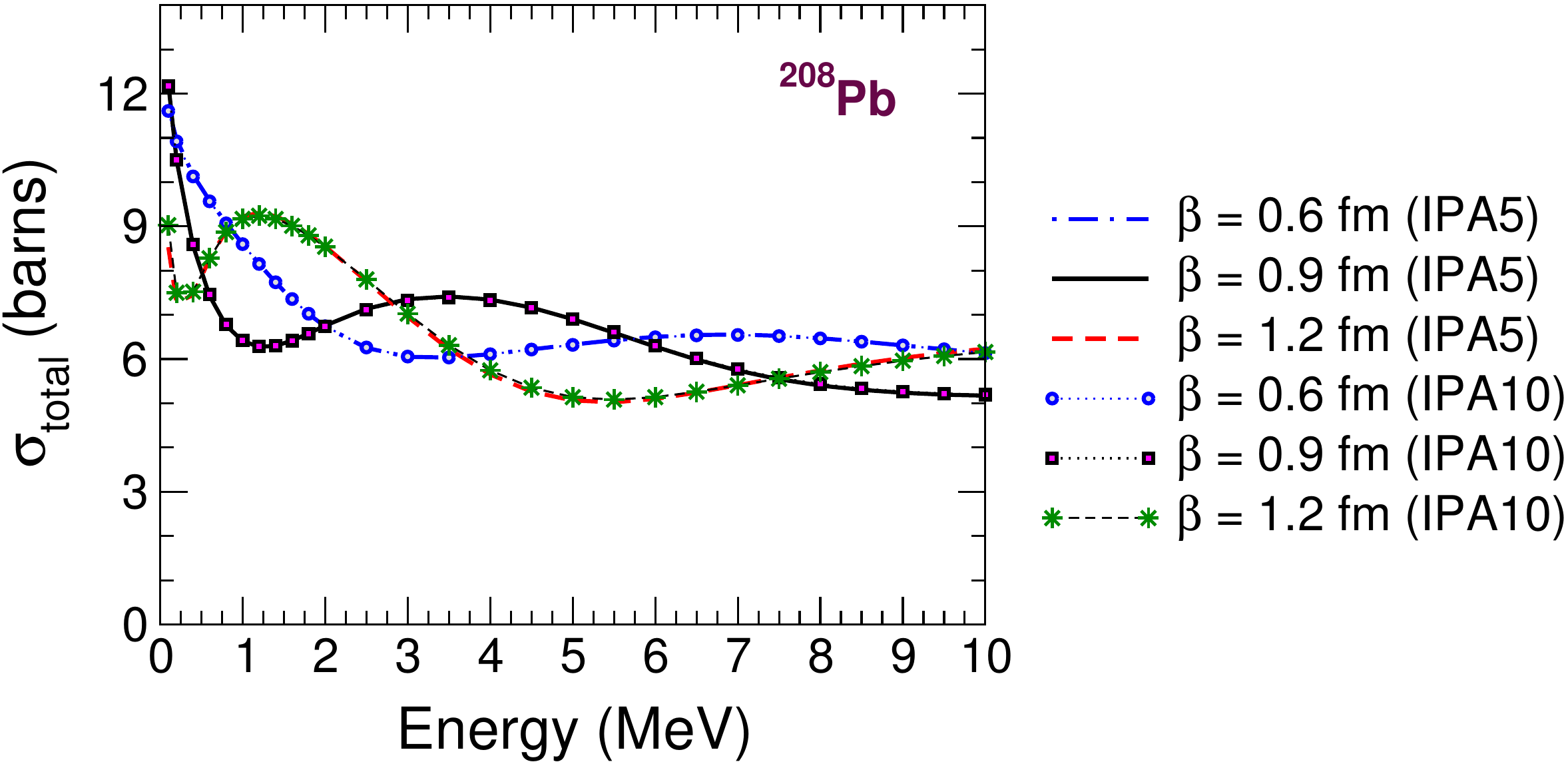}
\caption{Calculated total cross sections for neutron scattering off
$^{208}$Pb. Calculations are done using Gaussian form of nonlocality 
and TPM15 potential \cite{tpm15} for different values of $\beta$.
}
\label{f9}}
\end{figure*}

Next we explore the impact of different rms radius on Gaussian form
of nonlocality. For this we consider different values of $\beta$, 
namely, 0.6~fm, 0.9~fm and 1.2~fm in TPM15 potential \cite{tpm15} and 
calculate total cross sections using 5 iterations of IPA (IPA5). As 
an illustration, in Fig.~\ref{f9} we show the calculated cross sections
for neutron scattering off $^{208}$Pb. The cross sections are found to 
be extremely sensitive to $\beta$. However, it should be noted that 
$\beta$ is an additional parameter in TPM15 potential. Hence, in 
principle, any change in $\beta$ should be accompanied by refitting
of the potential parameters \cite{pb}. Nevertheless, this study illustrates 
the numerical robustness of IPA against the range of nonlocality.

In order to test the convergence properties of IPA, in Fig.~\ref{f9}
we also show the calculated cross sections with 10 iterations of IPA 
(labeled as IPA10) for different values of $\beta$. As can be seen,
irrespective of $\beta$ value, convergence is achieved with 5 
iterations. Further, we would like to point out that the runtime
required for IPA10 is only marginally longer than that for IPA5. 

Thus, it can be concluded that IPA is a robust technique and its
validity seems to be independent of the choice of nonlocal form factor.

\section{Summary and Conclusion}
A very efficient and highly precise technique to solve the integro-differential
equation appearing in scattering problem is developed. It is achieved by 
employing Taylor approximation to the radial wave function. This scheme
transforms the integro-differential equation to a second-order homogeneous
differential equation which can be solved easily. 

The observables obtained by the Taylor scheme for neutrons scattering 
off $^{24}$Mg, $^{40}$Ca, $^{100}$Mo and $^{208}$Pb are found to be 
within 8$\%$ of those obtained by the IMVT scheme at all the projectile
energies. We have demonstrated that the precision of solution can be 
improved further by using the ``Iterative Perturbative Approach", which
calculates the successive corrections to the solution obtained by using
the Taylor scheme. With just 5 iterations of IPA the observables for all
the cases and at all the energies are found to be within 2$\%$ of those
obtained by the IMVT scheme without any appreciable change in the run 
time. Further, the calculated observables are in accord with the 
experiments for all the cases.

The technique developed here is found to be robust and numerically stable. 
This conclusion seems to be independent of the choice of the form of 
nonlocality. Therefore, it is expected to be useful in diverse areas of 
science where existence of nonlocality leads to an integro-differential 
equation.

\begin{acknowledgments}
We thank B. K. Jain, Swagata Sarkar and R. C. Cowsik for their critical
feedback. NJU acknowledges financial support from SERB, Govt. of India 
(grant number YSS/2015/000900). AB acknowledges financial support from 
DST, Govt. of India (grant number DST/INT/SWD/VR/P-04/2014).
\end{acknowledgments}

\nocite{*}

\bibliographystyle{apsrev4-1}
\bibliography{paper}

\end{document}